# NOISE BASED LOGIC: WHY NOISE? A COMPARATIVE STUDY OF THE NECESSITY OF RANDOMNESS OUT OF ORTHOGONALITY

HE WEN[a, b], LASZLO B. KISH[a]

[a]*Texas A&M University, Department of Electrical and Computer Engineering, College Station, TX 77843-3128, USA*

[b]*Hunan University, College of Electrical and Information Engineering, Changsha, 410082, China*

**Abstract:** Although noise-based logic shows potential advantages of reduced power dissipation and the ability of large parallel operations with low hardware and time complexity the question still persist: is randomness really needed out of orthogonality? In this Letter, after some general thermodynamical considerations, we show relevant examples where we compare the computational complexity of logic systems based on orthogonal noise and sinusoidal signals, respectively. The conclusion is that in certain special-purpose applications noise-based logic is exponentially better than its sinusoidal version: its computational complexity can be exponentially smaller to perform the same task.

## 1. Introduction

Recently, new, non-conventional ways of deterministic (non-probabilistic) multi-valued *noise based logic* (*NBL*) system [1] has been introduced for lower energy consumption and high-complexity parallel operations in post-Moore-law-chips. These new methods are inspired by the fact that the brain uses noise and its statistical properties for information processing. Historically, *NBL* was the next logical step after stealth communications scheme [2], and the enhanced-Johnson-noise based information theoretically secure key distribution method [3], which were communication schemes where random noises carried the information bit.

The *NBL* uses electronic thermal noise as information carrier, as shown in Fig.1, where the logic 0 and 1 signals are represented by independent stochastic noise sources of zero mean (except the unipolar spike sequence based brain logic scheme (see below) where the mean is non-zero). For example, suppose $V_i(t)$ ($i$=1, ..., $N$) are independent random processes with zero mean, which will hold that $\langle V_i(t)V_j(t)\rangle = \delta_{i,j}$, where $\delta_{i,j}$ is the Kronecker symbol (i.e., for $i=j$, $\delta_{i,j}=1$, otherwise $\delta_{i,j}=0$). Note, the mathematically exact Kronecker values naturally assume infinite averaging times. In Fig.1, an orthogonal system of random noise processes forms the reference signal system (orthogonal base) of logic values.

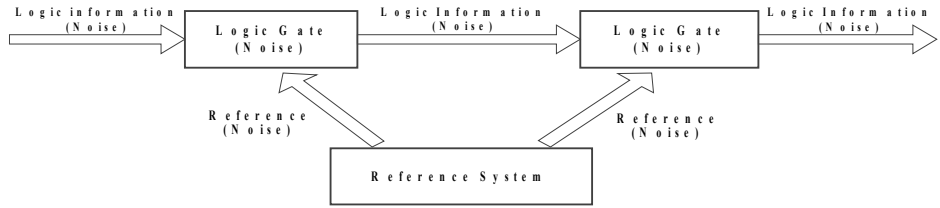

Fig.1 Basic structure of noise-based logic system

Within the *NBL* framework, *instantaneous noise-based logic* (*INBL*) [4, 5], has been introduced where the logic values are encoded into nonzero, bipolar, independent *random telegraph waves* (*RTW*). The *RTW*s used in the *INBL* are random square waves, which take the value of +1 or -1 with probability 0.5 at the beginning of each clock period and stay with this value during the rest of the clock duration. In the non-squeezed *INBL*, for the *r-th* noise bit, there are two reference *RTW*s, $H_r(t)$ and $L_r(t)$, representing its logic values, respectively.

Moreover, logic schemes based on other orthogonal processes, e.g. sinusoidal signals [6], have also been proposed to identify a signal without ambiguity. Thus, although the *NBL* has several potential advantages, such as reduced error propagation and power dissipation, it is still interesting to answer the question: why noise? Is randomness really needed out of orthogonality? In this letter, we compare the *RTW*s and sinusoidal signals to illustrate the necessity and importance of randomness when representing logical values with orthogonal signals. For the sake of simplicity, this paper will concentrate on the instantaneous logic scheme and special-purpose applications.

In the rest of the paper, we will compare *NB*L with an alternative orthogonal deterministic realizations using sinusoidal signals. It was already pointed out for universal, correlator-based *NBL* that sinusoidal representations are inferior [1] and a subsequent circuit-analysis-based study confirmed the negative expectations [7]. However no analysis or comparison exists between instantaneous *NBL* and its sinusoidal alternatives.

**2. General arguments: the case of entropy**

Preliminary considerations of the energy dissipation issue in biological informatics have been given in [8-10]. Noise-based logic was inspired by the stochastic neural signals in the brain [1] and a particular brain logic scheme utilizing stochastic signals was also proposed [9, 10]. The human brain consumes some 10-20 Watt power while it operates with noise (stochastic neural spike sequences) as information carrier. The natural questions occurs: is there a physical principle that favors stochastic processes for energy-friendly computation?

In this section, we consider the related aspects for noise-based logic by incorporating Brillouin's negentropy principle [11,12]. When energy dissipation is an issue, the problem of entropy can provide a generic answer why certain types of noises may serve better than deterministic signals. Let us imagine an isolated system in thermal equilibrium.

According to Brillouin's negentropy principle [11,12] if we set up an oscillator to generate a deterministic signal in a subvolume of an isolated thermodynamical system, that will unavoidably lead to entropy production, heat, emitted to the rest of the system. According to Brillouin's negentropy principle, the deterministic signal has negative entropy (negentropy) $-dS$ due to its reduced relative uncertainty. According to the second law of thermodynamics, the total entropy of the whole system cannot decrease, which means that the entropy in the rest of the system (out of the signal channel) must increase by $dS$. That means generating heat. Moreover, oscillators have loss, which means, to keep the oscillation going, a continuous energy injection is needed, which results in a steady heating power. Therefore, if we set up a system of $2N$ sinusoidal oscillators with different frequencies to provide orthogonal signals for the $N$-bits reference system, itself the realization of this signal system will require an upfront energy dissipation and a subsequent steady power dissipation.

On the other hand, spontaneous fluctuations of thermal origin (thermal noise) are present in the system. As a relevant example for noise-based logic, we can utilize these thermal fluctuations as reference signals. For example, we can use the thermal noise of $2N$ resistors as the orthogonal signals for the reference signal system of $N$ bits noise-

*Why noise?*

based logic. The realization of this signal system , at least over short distances where the loss is negligible, requires zero energy dissipation because these signals are granted by the existence of thermal fluctuations in thermal equilibrium.

Even though, the above considerations are only about the reference signal system, and computations would require energy dissipation even with noise, this simple example indicates that noise can have a key role when energy-friendly computation is necessary.

**3. Hardware complexity**

A general problem with today's binary logic is that, the DC voltage levels representing the different bits and bit values can be considered one-dimensional vectors [1]. These vectors are not orthogonal to each other thus, the classic binary logic can not perform high parallelism with low hardware complexity.

However, the *NBL* can represent a multidimensional vector system utilizing the orthogonality of independent noises [1]. Generally, 2*N* independent (orthogonal) noises form *N* noise-bits, where in a single noise bit *X*, one noise represents the *L* and another the *H* value of the bit, respectively. Products of the base vectors lead out from the base and form a *hyperspace* with $2^N$ dimensions [1,13]. A logic signal that propagates in a single wire can be constructed by the binary superposition (on or off) of hyperspace vectors, which results in $2^{2^N}$ different logic values [2]. Consequently, *N* noise-bits correspond to $2^N$ classical bits in a single wire [13].

For applications mimicking quantum computing, the *N*-bit long hyperspace vectors (product states), $W=X_1X_2...X_N$ and their superpositions have key importance [13], where these product states represent the basis vectors of the $2^N$ dimensional Hilbert space of quantum informatics. Then, there are $2^N$ different products in an *N*-bit logic system. Thus, in a classical computer or utilizing the sinusoidal signals, the corresponding $2^N$ dimensional *universe,* which is the uniform superposition of all the *N*-bit long product states has complexity of $O(N*2^N)$ [13].

For applications, it is essential that the universe can be synthesized with low-complexity operations. With the Achilles heel method, it can be generated [13] with the following operation that has only $O(2N)$ complexity

$$Y(2^N) = \prod_{r=1}^{N} \left(L_r + H_r\right) \qquad (2)$$

The easy synthesis of the universe and the resulting high parallelism when manipulating it with simple, low-complexity operations, is very attractive. For example, when the number of noise-bits $N=200$, a proper algebraic operation acting on this superposition in a single wire can potentially achieve a (special-purpose) parallel operation that would require a $200*2^{200}$ bits operation in classical computing [14].

**4. Bandwidth (time complexity) of product strings**

As it is described above, to set up a *hyperspace* with $2^N$ dimensions, $2N$ independent (orthogonal) noises are required to form $N$ noise-bits. Similarly, sinusoidal signals can also be used to represent the noise-bit values [1,7]. In this section, we compare the time complexity of RTW hyperspace based logic with the corresponding sinusoidal one. It will be shown that the required bandwidth of the sinus-based product string (hyperspace vector) will scale with $2^N$ to avoid the *degeneracy* of the system.

Time complexity in these special-purpose applications has two sides:

*a*) The time complexity needed to set up the hyperspace vectors.

*b*) The time complexity needed to analyze/decode the result which, in the practically important cases, is a single product string because, while large-scale parallel calculations are being executed on the superposition of hyperspace vectors; the final result, on which the actual measurement is done, is typically a single vector.

*4.1 Time complexity needed to set up the hyperspace vectors*

Using RTWs to represent the bit values results in no time complexity increase when setting up a hyperspace vector, because the RTW representing the hyperspace vector has the same clock frequency as its components.

Concerning the sinusoidal signal system, first let us suppose that we fill up the "harmonics space" linearly with the bit values; for example, the subsequent odd harmonics represent the subsequent Low bit values and the subsequent even harmonics the subsequent High values (*linear frequency representation,* see Table I). Thus to represent $N$ "sinus-based" bits, $2N$ different harmonics are needed. In this case, we can define the logic values of the *r*-th bit as

*Why noise?*

$$L_r(t) = e^{j2\pi(2r-1)f_0 t} \quad \text{and} \quad H_r(t) = e^{j2\pi 2 r f_0 t} \quad (3)$$

In this case, the highest frequency in the system will be the sum of all the utilized harmonics from 1 to 2*N*, that is, $N(2N+1)f_0$, which is scaling with $N^2$. However, this system will be *degenerated* because of the frequency-overlapping when forming the hyperspace vector. For example, the product $L_1H_2$ and $H_1L_2$ will produce the same harmonics. This kind of degeneracy will cause significant information loss and make the system virtually useless.

Second, let us suppose that we fill up the "harmonics space" exponentially (*exponential frequency representation*) where the frequencies of harmonics form a 2-based geometric sequence. Then the subsequent bit values are represented by harmonics that scale with the power of 2 (see also Table 1):

$$L_r(t) = e^{j2\pi 2^{2r-2} f_0 t} \quad \text{and} \quad H_r(t) = e^{j2\pi 2^{2r-1} f_0 t} \quad (4)$$

As a result, the highest frequency in the system will be the sum of all the utilized harmonics which is $(2^{2N}-1)f_0$ that is scaling with $2^{2N}$. In conclusion to set up a single hyperspace vector of a sinus-based system requires exponential time complexity $O(2^{2N})$.

Table I. The logic values and corresponding frequencies for both the linear and exponential frequency representation

| Bit | Logic Value | Frequency | |
|---|---|---|---|
| | | Linear Representation | Exponential Representation |
| 1st | $L_1$ | $f_0$ | $f_0$ |
| | $H_1$ | $2f_0$ | $2f_0$ |
| 2nd | $L_2$ | $3f_0$ | $4f_0$ |
| | $H_2$ | $4f_0$ | $8f_0$ |
| ... | ... | ... | ... |
| *N*th | $L_N$ | $(2N-1)f_0$ | $2^{2N-2}f_0$ |
| | $H_N$ | $2Nf_0$ | $2^{2N-1}f_0$ |

*4.2 Time complexity needed to analyze/decode the result (hyperspace vector)*

To read out the bit values of this hyperspace vector requires a Fourier analysis and a time window of $1/f_0$. However, this time window contains frequency components as high as $(2^{2N}-1)f_0$, which means that the time complexity of the read-out operation is also $O(2^{2N})$.

Concerning the reading out the noise-bit values in a hyperspace vector Stacho [15] has developed an algorithm that provides an exponential speedup compared to a "brute force" determination. Stacho's algorithm provides that, when utilizing less than $t_\epsilon = N(\log_2 N)^{(1+\epsilon)}$ clock periods, the probability $P$ that the measurement algorithm fails is scaling as $P \propto 2^{-N}$ in the limit of large $N$, where for $N \to \infty$, $\epsilon \to 0$.

In time-shifted *NBL*, there is a polynomial (linear in $2N$) time complexity in setting up the hyperspace vector and also the same $O(N)$ complexity to read out the values, specifically, $2N * \log_4(N/P)$ of time steps are needed, which in this system means $\log_4(N/P)$ clock periods [14].

## 5. Conclusions

Our goal was to answer the following question: *Does noise-based logic really need "noise" (stochasticity) or an orthogonal system of sinusoidal signals is enough with similar computational complexity*?

Then answer is that "noise" must be an essential component of these special-purpose, orthogonal-signal-based logic systems for various reasons including low power dissipation (low entropy production) and low computational complexity.

**Acknowledgments**

Discussions with Sergey Bezrukov, Andreas Klappenecker and Laszlo Stacho are appreciated. This work has partially been supported by the National Natural Science Foundation of China under grant 61002035.

*Why noise?*

**References**

[1]     L.B. Kish, "Noise-based logic: Binary, multi-valued, or fuzzy, with optional superposition of logic states", *Physics Letters A* **373** (2009) 911-918.

[2]     L.B. Kish, "Stealth communication: Zero-power classical communication, zero-quantum quantum communication and environmental-noise communication", *Applied physics letters,* **87** (2005) 234109.

[3]     L.B. Kish, "Totally secure classical communication utilizing Johnson (-like) noise and Kirchoff's law", *Physics Letters A,* **352** (2006) 178-182.

[4]     L.B. Kish, S. Khatri, F. Peper, "Instantaneous noise-based logic", *Fluctuation and Noise Letters* **9** (2010) 323-330.

[5]     F. Peper, L. B. Kish, "Instantaneous, non-squeezed, noise-based logic", *Fluctuation and Noise Letters* **10** (2011) 231-237.

[6]     M. Ueda, M. Ueda, H. Takagi, M. J. Sato, T. Yanagida, I. Yamashita and K.Setsune, "Biologically-inspired stochastic vector matching for noise-robust information processing", *Physica A: Statistical Mechanics and its Applications,* **387** (2008) 4475-4481.

[7]     K.C. Bollapalli, S.P. Khatri, L.B. Kish, "Implementing digital logic with sinusoidal supplies [C]. Proceedings of the Design", *Automation & Test in Europe Conference & Exhibition (DATE)*, 2010: 315-318.

[8]     S.M. Bezrukov, L. B. Kish, "How much power does neural signal propagation need?", *Smart Materials and Structures,* **11** (2002) 800.

[9]     S. M. Bezrukov, L. B. Kish, "Deterministic multivalued logic scheme for information processing and routing in the brain", *Physics Letters A* **373** (2009) 2338-2342.

[10]    Z. Gingl, S. Khatri, L.B. Kish, "Towards brain-inspired computing", *Fluctuation and Noise Letters* **9** (2010) 403-412.

[11]    L. Brillouin Scientific Uncertainty and Information, (Academic Press, New York, 1964).

[12]    L. Brillouin, Science and Information Theory, (Academic Press, New York, 1962).

[13]    L.B. Kish, S. Khatri, S. Sethuraman, "Noise-based logic hyperspace with the superposition of 2(N) states in a single wire", *Physics Letters A* **373** (2009) 1928-1934.

[14]    H. Wen, L.B. Kish, A. Klappenecker, F. Peper, "New noise-based logic representations to avoid some problems with time complexity", *Fluctuation and Noise Letters*, accepted for publication (March 2012); http://arxiv.org/abs/1111.3859 .

[15]    L. Stacho, "Fast measurements of hyperspace vectors in noise-based logic", *Fluctuation and Noise Letters*, in press.